\input{aipcheck}

\documentclass[
    ,final            % use final for the camera ready runs
  ]
  {aipproc}

\layoutstyle{6x9}

\def\de{\delta}

\def\frac#1#2{{\textstyle{{#1}\over {#2}}}}

\def\lsim{\mathrel{\rlap{\lower4pt\hbox{\hskip1pt$\sim$}}
    \raise1pt\hbox{$<$}}}
\def\gsim{\mathrel{\rlap{\lower4pt\hbox{\hskip1pt$\sim$}}
    \raise1pt\hbox{$>$}}}
\def\sqr#1#2{{\vcenter{\vbox{\hrule height.#2pt
         \hbox{\vrule width.#2pt height#1pt \kern#1pt
         \vrule width.#2pt}
         \hrule height.#2pt}}}}

\def\lrpartial{\raise 1pt\hbox{$\stackrel\leftrightarrow\partial$}}

\def\etal{{\it et al.}}

\newcommand{\beq}{\begin{equation}}
\newcommand{\eeq}{\end{equation}}
\newcommand{\bea}{\begin{eqnarray}}
\newcommand{\eea}{\end{eqnarray}}
\newcommand{\rf}[1]{(\ref{#1})}

\begin{document}

{}\hfill MIT-CPT-3822

\title{Quantum-gravity phenomenology,\\ 
Lorentz symmetry, and the SME\footnote{Invited talk presented at SILAFAE '06, Puerto Vallarta, Mexico.}}


\author{Ralf Lehnert}{
  address={Center for Theoretical Physics,
Massachusetts Institute of Technology, Cambridge, MA 02139}
}

\begin{abstract}
 Violations of spacetime symmetries 
 have recently been identified as promising signatures for physics underlying the Standard Model. 
 The present talk gives an overview over various topics in this field: 
 The motivations for spacetime-symmetry research, 
 including some mechanisms for Lorentz breaking, 
 are reviewed. 
 An effective field theory called the Standard-Model Extension (SME) 
 for the description of the resulting low-energy effects 
 is introduced, 
 and some experimental tests of Lorentz and CPT invariance 
 are listed.
\end{abstract}

\maketitle

%%%%%%%%%%%%%%%%%%%%%%%%%%%%%%%%%%%%%%%%%%%%
%% MAINMATTER
%%%%%%%%%%%%%%%%%%%%%%%%%%%%%%%%%%%%%%%%%%%%

\section{Introduction}
\label{intro} 

Perhaps the most challenging open question 
in present-day fundamental physics 
concerns a unified quantum theory 
of all interactions including gravity. 
To date, 
a tremendous amount of theoretical work 
has been devoted to various approaches 
addressing this question. 
However, 
experimental efforts in this line of research 
are hampered, 
primarily because of the expected Planck suppression 
of the corresponding effects 
at presently attainable energies. 
A possible avenue to circumvent this issue 
is to scan the predictions of a given candidate underlying theory 
for effects 
that could be present already at lower energy scales. 
For example, 
one can search for novel particles, such as those required by supersymmetry, 
or large extra dimension. 

Another promising approach is to ask 
which type of effect {\em can} be measured with Planck precision, 
and then determine whether such effects are indeed allowed 
in theoretical approaches to a more fundamental theory. 
In this context, 
tests of invariance properties appear to be an excellent candidate: 
symmetries allow {\em exact} theoretical predictions, 
and they are typically amenable ultrahigh-precision experiments. 
From a quantum-gravity perspective, 
spacetime symmetries could be particularly promising: 
gravity governs the dynamics of space and time, 
so that a quantum theory of gravitation 
is likely to affect the structure of spacetime. 
For example, 
typical underlying models 
can involve more than four dimensions, 
operator-valued noncommuting coordinates, 
or a certain discreteness of space. 

The present talk further explores the idea 
that spacetime symmetries---and in particular Lorentz invariance---may be violated 
by small amounts 
as a result of more fundamental physics. 
We begin by studying the interplay between various spacetime symmetries.  
Then, 
two sample mechanisms for Lorentz and CPT breakdown 
in Lorentz-symmetric approaches to underlying physics  
are reviewed. 
The last section  
recounts the basic philosophy 
behind the construction of the Standard-Model Extension (SME), 
which provides the modern effective-field-theory framework 
for describing Lorentz and CPT violation. 
This section also lists 
some tests of Lorentz and CPT symmetry.

\section{The interplay between various spacetime-symmetry violations} 

Spacetime transformations 
that, to the best of our knowledge, 
are associated with a symmetry of nature 
are closely intertwined in the Poincar\'e group. 
Consider the case 
in which symmetry is lost under one (or more) transformations. 
The question arises 
as to whether the remaining transformations 
still determine an invariance,  
or whether the violation of one set of spacetime symmetries 
typically leads to the breakdown of other invariances contained in the Poincar\'e group. 
The purpose of the present section is to explore this question. 

Let us first consider the case of a discrete transformation---the 
combination of charge conjugation (C), parity inversion (P), and time reversal (T). 
The famous CPT theorem \cite{cpt} states roughly 
that, under a few mild assumptions, 
CPT invariance is a consequence of  
conventional quantum mechanics and Lorentz symmetry. 
Thus, 
when CPT invariance is broken, 
one or more ingredients of the CPT theorem 
can no longer hold true. 
The questions arises 
which one of these ingredients should be abandoned. 
Both the Lorentz and the CPT transformations involve spacetime, 
which suggests that CPT breaking implies Lorentz violation. 
This result has been proved in the framework of axiomatic field theory by Greenberg \cite{green02}. 
This ``anti-CPT theorem'' 
essentially  asserts 
that in local, unitary, relativistic point-particle quantum field theories
a breakdown of CPT symmetry always comes with Lorentz violation. 
Note, 
however, 
that the converse of this result, 
i.e., Lorentz violation implies CPT breakdown, 
is false in general. 
We see that as a corollary, 
CPT tests are at the same time Lorentz tests. 
We finally note 
that other types of CPT breaking 
would require further deviations from conventional physics, 
such as unconventional quantum mechanics \cite{mav}.

We next look at a situation 
in which translational invariance is violated. 
(A mechanism for this effect is studied in a subsequent section of this talk.) 
In this case, 
the energy--momentum tensor $\theta^{\mu\nu}$, 
which generates translations, 
is usually no longer conserved.
To see 
that this also affects Lorentz symmetry, 
we consider the angular-momentum tensor $J^{\mu\nu}$, 
given by
\begin{equation} 
J^{\mu\nu}=\int d^3x \;\big(\theta^{0\mu}x^{\nu}-\theta^{0\nu}x^{\mu}\big). 
\label{gen} 
\end{equation}
In the Poincar\'e-symmetric case, 
this tensor is the generator of Lorentz transformations. 
Note 
that energy--momentum tensor $\theta^{\mu\nu}$ 
appears in the definition of $J^{\mu\nu}$. 
Since $\theta^{\mu\nu}$ is taken as not conserved in the present case, 
$J^{\mu\nu}$ 
will usually depend on time. 
Then, 
the expression \rf{gen} 
no longer determines the conventional time-independent 
Lorentz-transformation generators. 
In fact, 
such generators will typically no longer exist, 
and then Lorentz invariance ceases to be assured. 
In other words, 
translation-symmetry breaking 
is in such cases associated with Lorentz violation.

\section{Examples of mechanisms for spacetime-symmetry violation} 
\label{mechanisms} 

The preceding section has shown 
that the breakdown of certain subsets of spacetime symmetries 
can result in the violation of additional spacetime invariances. 
But the question remains 
how spacetime symmetries can be broken 
in a Poincar\'e-invariant underlying theory in the first place. 
Various mechanisms for symmetry violations  
have been devised in a number of approaches to fundamental physics, 
such as 
strings \cite{kps}, 
spacetime foam \cite{sf}, 
noncommutative geometry \cite{noncom},
nontrivial spacetime topology \cite{klink}, 
and cosmologically varying scalars \cite{varscal}. 
The present section 
gives a somewhat more detailed description 
of two of the above mechanisms---spontaneous Lorentz and CPT breaking in string theory 
and Lorentz and CPT violation through spacetime-dependent scalars.

{\bf Spontaneous Lorentz and CPT violation.} 
Spontaneous breakdown of Lorentz and CPT symmetry 
can occur in the context of the field theory 
of the open bosonic string \cite{kps}. 
The mechanism of spontaneous symmetry violation 
is quite attractive from a theoretical viewpoint 
because the dynamics remains invariant 
under the symmetry in question.
It is the ground-state solution of the system 
that fails to exhibit all invariances of the Hamiltonian, 
and is therefore said to break the corresponding symmetries. 
Instances of spontaneous symmetry violation 
can readily be identified 
in solid-state physics, 
in the physics of elastic media, 
and in elementary-particle theory. 

\begin{figure}
	\label{fig3}
  \includegraphics[height=.8\textheight]{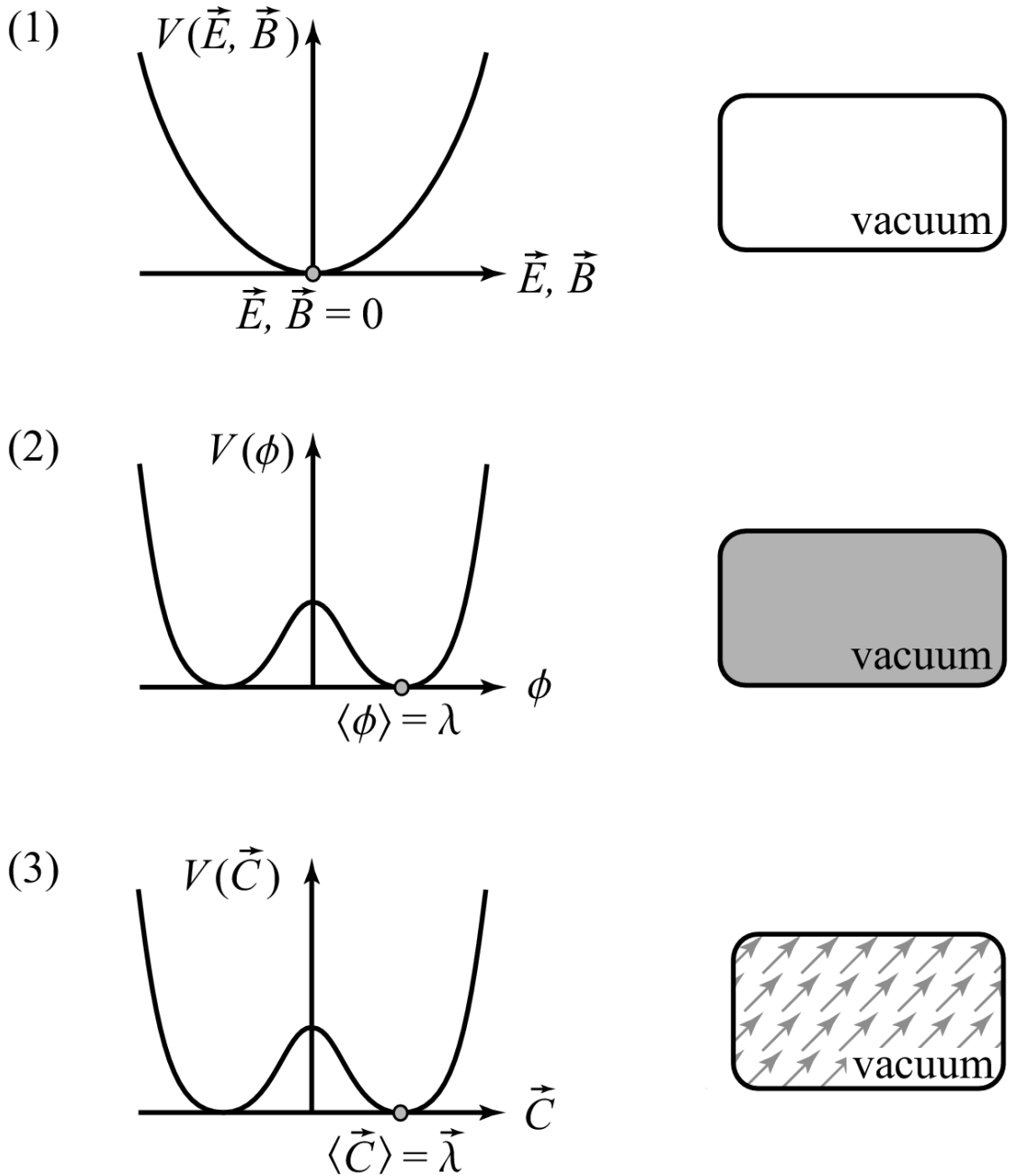}
  \caption{
  Spontaneous symmetry breaking. 
In Maxwell's classical electromagnetism (1), 
the state with the lowest energy is attained 
for $\vec{E}=0$ and $\vec{B}=0$. 
The classical Maxwell vacuum is therefore empty. 
The Higgs-type field (2) 
possesses interactions 
with an energy density $V(\phi)$ 
that triggers a non-zero value 
of $\phi$ in the ground state. 
The vacuum contains a scalar condensate, 
which is shown in gray. 
Lorentz and CPT symmetry are still intact. 
(However, other, internal symmetries may be broken.) 
Vector fields occurring, for instance, 
in string field theory (3) 
can have energy densities  
analogous to those of the Higgs field. 
Such interactions would lead to a nonvanishing field value 
in the lowest-energy configuration. 
The resulting VEV of such a vector field 
causes a special direction in the vacuum, 
which breaks Lorentz and possibly CPT invariance.
  }
\end{figure}

In what follows, 
we will discuss three sample physical systems. 
The features of these examples 
will enable us 
to gain further intuition 
about spontaneous Lorentz and CPT breakdown
in a step-by-step fashion. 
An illustration 
that supports these three examples 
is given in Fig.\ \ref{fig3}.

Our first system is classical electrodynamics. 
Within this context, 
the energy density $V(\vec{E},\vec{B})$ 
of any pattern of electric and magnetic fields $\vec{E}$ and $\vec{B}$, respectively,
is determined by  
\begin{equation}
\label{max_en_den}
V(\vec{E},\vec{B})=\frac{1}{2} \left(\vec{E}^2+\vec{B}^2\right)\, ,
\end{equation}
where we have employed natural units. 
Given any solution  of the classical Maxwell equations, 
Eq.\ (\ref{max_en_den}) 
yields the energy stored in the electromagnetic fields. 
Any field configuration involving a nonzero field 
is associated with a strictly positive energy. 
Only when both $\vec{E}$ and $\vec{B}$ are zero everywhere, 
we have vanishing field energy. 
The lowest-energy configuration of a system 
is usually identified with its ground state or vacuum.
It is thus apparent 
that in classical electrodynamics 
the vacuum contains no fields; 
the Maxwell vacuum is empty. 

The second sample system we examine 
is a Higgs-type field $\phi$. 
Unlike the electromagnetic fields considered above, 
the Higgs field is a scalar. 
Scalar fields of this type 
are believed to occur in Nature;
they are, in fact, part of the Standard Model of particle physics. 
As before, 
we consider the energy density of $\phi$. 
In the absence of spacetime dependence, 
i.e., $\partial_{\mu}\phi=0$, 
the energy density $V(\phi)$ obeys
\begin{equation} 
\label{scal_en_den} 
V(\phi)=(\phi^2-\lambda^2)^2\, , 
\end{equation} 
where $\lambda$ is a constant. 
A possible spacetime dependence of $\phi$ 
would add (positive definite) kinetic-type energy to this expression, 
which is uninteresting in the present context. 
We therefore see that the lowest possible energy state of $\phi$ 
possesses zero energy. 
As opposed to the electrodynamics case, 
this state is attained for {\em finite} field values $\phi=\pm\lambda$. 
It is thus apparent 
that in the presence of such Higgs-type fields 
the vacuum is filled with a condensate
$\phi_{vac}\equiv\langle\phi\rangle=\pm\lambda$. 
In quantum theory, 
the condensate $\langle\phi\rangle$ 
is referred to as the vacuum expectation value (VEV) 
of $\phi$. 
One of the physical effects 
associated with the VEV of the Standard-Model Higgs 
is to cause mass terms for many elementary particles. 
It is important to note 
that $\langle\phi\rangle$ is a scalar, 
so it does {\it not} select a Lorentz-violating spacetime direction.

The third sample system we consider 
concerns a (hypothetical) vector field $\vec{C}$. 
Such a vector field clearly lacks observational evidence.
However, 
such a field may be contained in candidate fundamental theories, 
such as strings. 
We focus here on the non-relativistic physics, 
which is sufficient to illustrate rotation violation; 
a relativistic generalization can easily be obtained. 
Paralleling the previous Higgs-field case, 
we take the potential-energy density 
for $\vec{C}$ to be given by
\begin{equation} 
\label{vec_en_den} 
V(\vec{C})=(\vec{C}^2-\lambda^2)^2\, . 
\end{equation} 
Spacetime dependence of $\vec{C}$ would contribute 
positive-definite kinetic-energy contributions. 
Equation \rf{vec_en_den} shows 
that $V=0$ is the lowest energy for the system. 
As for the Higgs field, 
this lowest-energy state 
requires $\vec{C}$ to be nonvanishing: 
$\vec{C}_{vac}\equiv\langle\vec{C}\rangle=\vec{\lambda}$. 
Here, 
$\vec{\lambda}$ is any spacetime-constant vector 
that obeys $\vec{\lambda}^2=\lambda^2$. 
As before, 
the vacuum fills with the VEV of the field $\vec{C}$ 
and is therefore not empty. 
Since $\langle\vec{C}\rangle=\vec{\lambda}$ is a constant vector, 
the vacuum contains an intrinsic direction, 
which violates rotational invariance 
and therefore also Lorentz symmetry.

{\bf Cosmologically varying scalars.} 
As noted at the beginning of this section,
a scalar varying on cosmological scales 
leads to Lorentz violation. 
This can be established 
with our result from the previous section 
that the breakdown of translational invariance 
(here via the spacetime dependence of the scalar)  
typically leads to the loss of Lorentz symmetry. 
This effect is independent of the mechanism driving the variation of the scalar. 
The remainder of this section 
gives a more detailed discussion of this result 
with the goal of providing further intuition. 

We begin by establishing the effect at the Lagrangian level. 
To this end, 
consider two scalar fields $\phi$ and $\Phi$ 
and a varying coupling $\xi(x)$. 
It is the spacetime dependence of $\xi(x)$ 
that will lead to Lorentz violation;
$\phi$ and $\Phi$ are sample dynamical variables 
that could in principle be replaced by vector or spinor fields. 
Let the Lagrangian $\mathcal{L}$ of the system contain a kinetic-type term 
of the form $\xi(x)\,\partial^{\mu}\phi\,\partial_{\mu}\Phi$.
A suitable integration by parts 
at the level of the action 
will produce a boundary term 
that can be dropped 
while leaving unaffected the equations of motion. 
The resulting Lagrangian $\mathcal{L}'$ 
will then contain a term of the form 
\begin{equation}
\mathcal{L}'\supset -K^{\mu}\phi\,\partial_{\mu}\Phi\, .
\label{example1}
\end{equation}
Here, 
the vector $K^{\mu}\equiv\partial^{\mu}\xi$ is an external
nondynamical 4-vector. 
Such a quantity breaks Lorentz symmetry 
because it selects a preferred direction in spacetime.  
If $\xi$ varies on cosmological scales, 
$K^{\mu}$ is essentially a constant with respect to local physics, 
such as solar-system physics. 

\begin{figure}
	\label{fig4}
  \includegraphics[height=.18\textheight]{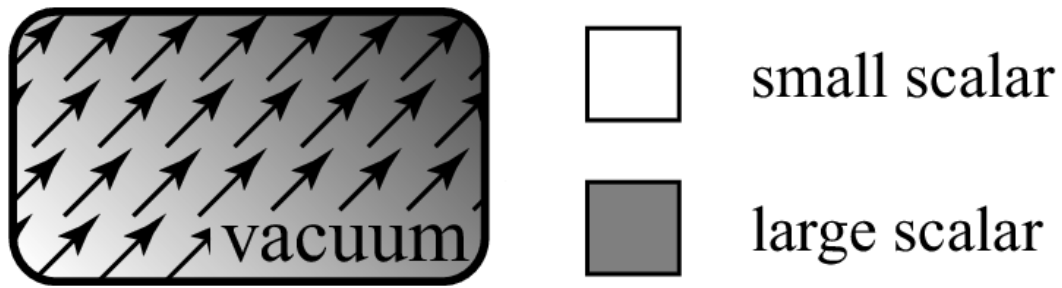}
  \caption{
  Lorentz breakdown via spacetime-dependent scalars.
The shade of gray displays the size of the scalar. 
Lighter areas correspond to smaller values 
and darker areas to larger values.
The black arrows symbolize the gradient of the scalar. 
This gradient selects a preferred direction in spacetime,  
so that Lorentz symmetry is violated. 
  }
\end{figure}

To gain further understanding 
of the Lorentz breakdown resulting 
from a varying scalar 
consider the following intuitive picture. 
The variation of the scalar implies 
that there is some region 
in which its 4-gradient is nonzero. 
Such a 4-gradient 
is associated with a preferred direction in spacetime. 
This is illustrated in Fig.\ \ref{fig4}. 
For instance, 
consider a particle 
that possesses interactions with this background scalar. 
When the motion of such a particle is along the gradient of the scalar, 
its propagation features may be different from the situation 
in which this particle moves perpendicular to the gradient. 
Since physically inequivalent directions 
correspond to anisotropies, 
rotational invariance, 
and thus Lorentz symmetry, 
must be violated. 

\section{The Standard-Model Extension}
\label{smesec}

Once Lorentz and CPT violation is identified 
as a possible signature 
for physics underlying the Standard Model 
(and possibly arising at the Planck scale), 
it is desirable to have at ones disposal 
a test framework 
for the description of the resulting low-energy effects. 
Perhaps the most important use of such  
a test framework is the identification of suitable Lorentz tests. 
Another important advantage of a test model is provided by the fact 
that it can be used to compare different Lorentz and CPT tests 
if the model is broad enough. 
The consistency of the low-energy framework 
may also put some theoretical constraints 
on possible high-energy underlying models. 

A test model for a certain high-energy underlying theory 
can usually be obtained by taking the low-energy limit of that theory. 
In the present case of Lorentz and CPT violation 
we do not follow this approach for two reasons. 
First, 
no complete and realistic fundamental theory is currently known. 
It is therefore desirable to have test framework 
for Lorentz and CPT symmetry 
that is relatively independent of the details of candidate underlying models. 
This permits a comprehensive search for Lorentz and CPT breakdown. 
Second, 
in some approaches to quantum gravity 
the physical vacuum state is currently unknown. 
For those models, 
the standard determination of a unique low-energy limit therefore fails. 
In light of these facts, 
we must rather proceed by constructing a test framework 
of sufficient generality to include the largest class of possible Lorentz and CPT violations
consistent with certain more fundamental principles. 

In what follows, 
we review the construction of the flat-spacetime limit of the Standard-Model Extension (SME),
which is the modern framework for the description of the low-energy effects 
of Lorentz and CPT breakdown \cite{sme}.\footnote{A similar methodology can also be applied 
in curved-spacetime situations involving gravity \cite{grav}.} 
The basic idea is to start with the Standard-Model Lagrangian ${\mathcal L}_{\rm SM}$
and add terms that violate Lorentz and possibly also CPT symmetry: 
\begin{equation} 
{\mathcal L}_{\rm SME}={\mathcal L}_{\rm SM}+\de {\mathcal L}\; . 
\label{sme} 
\end{equation}
Here, 
the SME Lagrangian is denoted by ${\mathcal L}_{\rm SME}$  
and Lorentz- and CPT-breaking corrections are collected in $\de {\mathcal L}$. 
As discussed above, 
we need to construct $\de {\mathcal L}$ by hand. 
From the arguments in the previous sections 
we know 
that the symmetry-violating effects appear as background vectors or tensors in the vacuum. 
To be observable, 
they must couple to conventional fields. 
Because we insist on the fundamental principle of coordinate independence, 
this coupling must be a covariant contraction, 
so that $\de {\mathcal L}$ transforms as a scalar under changes of the (inertial) coordinate system. 
For example, 
\begin{equation} 
\de {\mathcal L}\supset b^{\mu}\,\overline{\psi}\gamma_5\gamma_{\mu}\psi\; . 
\label{example} 
\end{equation}
Here, 
$b^{\mu}$ is a Lorentz- and CPT-violating background 
assumed to be generated in some underlying theory. 
It is a free coefficient, 
which can be searched for in suitable tests. 
Clearly, 
$b^{\mu}$ must be extremely small 
on observational grounds. 
The quantity $\overline{\psi}\gamma_5\gamma_{\mu}\psi$ 
denotes the usual chiral current of a Standard-Model fermion. 
Note that all Lorentz indices are properly contracted,  
so that $b^{\mu}\,\overline{\psi}\gamma_5\gamma_{\mu}\psi$ 
is a coordinate scalar.

Clearly, 
this construction yields infinitely many contributions to $\de {\mathcal L}$, 
most of which would be expected to be subleading. 
For phenomenological purposes, 
it therefore seems practical and justified 
to consider only a subset of contributions to $\de {\mathcal L}$ 
that satisfies certain additional requirements. 
For example, power-counting renormalizability, 
translational invariance, 
and the usual gauge symmetries 
are commonly imposed.
This ``minimal SME''
has provided the basis 
for a number of 
investigations 
of Lorentz and CPT breakdown
involving 
mesons \cite{kexpt},
baryons \cite{ccexpt,spaceexpt,cane},
electrons \cite{eexpt,eexpt2,eexpt3},
photons \cite{photon},
muons \cite{muons}, 
and the Higgs sector \cite{higgs}. 
An analysis involving the gravity sector 
has recently also been performed \cite{grav}.
We note 
that neutrino-oscillation experiments
offer discovery potential \cite{sme,neutrinos}.

%%%%%%%%%%%%%%%%%%%%%%%%%%%%%%%%%%%%%%%%%%%%%%%%
%% BACKMATTER
%%%%%%%%%%%%%%%%%%%%%%%%%%%%%%%%%%%%%%%%%%%%%%%%

\begin{theacknowledgments}
The author would like to thank the organizers 
for the invitation to this stimulating meeting 
and for financial support. 
This work was also supported 
by the U.S.\ Department of Energy 
under cooperative research agreement No.\ DE-FG02-05ER41360 
and by the European Commission 
under Grant No.\ MOIF-CT-2005-008687. 
\end{theacknowledgments}

\end{document}